\documentclass[floats,floatfix,showpacs,preprintnumbers,amssymb,prd,twocolumn,superscriptaddress,nofootinbib,nolongbibliography,reprint]{revtex4-2}

\usepackage{amssymb,amsmath,verbatim,mathtools,needspace,enumitem,etoolbox,graphicx,physics,microtype,afterpage,xspace,tabularx,lmodern,multirow}
\usepackage{gensymb}
\usepackage[normalem]{ulem}
\usepackage[dvipsnames, usenames]{xcolor}
\usepackage{xr-hyper}
\definecolor{linkcolor}{rgb}{0.0,0.3,0.5}
\usepackage[unicode, colorlinks=true, linkcolor=linkcolor, citecolor=linkcolor, filecolor=linkcolor, urlcolor=linkcolor, linktocpage, breaklinks]{hyperref}
\usepackage[all]{hypcap}
\usepackage[T1]{fontenc}
\usepackage[utf8]{inputenc}
\usepackage[usenames,dvipsnames]{xcolor}
\hypersetup{colorlinks=true,citecolor=romared,linkcolor=romared,urlcolor=romared}

\definecolor{romared}{RGB}{142,0,28}

\newcommand{\be}{\begin{equation}}
\newcommand{\ee}{\end{equation}}

\def\be{\begin{equation}}
\def\ee{\end{equation}}
\newcommand{\beq}{\begin{eqnarray}}
\newcommand{\eeq}{\end{eqnarray}}

\usepackage{makecell}
\usepackage{soul}

\newcolumntype{Y}{>{\centering\arraybackslash}X}

\makeatletter
\newcommand*{\addFileDependency}[1]{
  \typeout{(#1)}
  \@addtofilelist{#1}
  \IfFileExists{#1}{}{\typeout{No file #1.}}
}
\makeatother

\newcommand*{\myexternaldocument}[1]{%
    \externaldocument{#1}%
    \addFileDependency{#1.tex}%
    \addFileDependency{#1.aux}%
}

\myexternaldocument{Supp}

\begin{document}
\title{Constraints on massive gravity from dipolar mode excitations}

\author{Vitor Cardoso} 
\affiliation{Niels Bohr International Academy, Niels Bohr Institute, Blegdamsvej 17, 2100 Copenhagen, Denmark}
\affiliation{CENTRA, Departamento de F\'{\i}sica, Instituto Superior T\'ecnico -- IST, Universidade de Lisboa -- UL, Avenida Rovisco Pais 1, 1049-001 Lisboa, Portugal}
\author{Francisco Duque}
\affiliation{CENTRA, Departamento de F\'{\i}sica, Instituto Superior T\'ecnico -- IST, Universidade de Lisboa -- UL, Avenida Rovisco Pais 1, 1049-001 Lisboa, Portugal}
\author{Andrea Maselli}
\affiliation{Gran Sasso Science Institute (GSSI), I-67100 L’Aquila, Italy}
\affiliation{INFN, Laboratori Nazionali del Gran Sasso, I-67100 Assergi, Italy}
\author{David Pere\~niguez}
\affiliation{Niels Bohr International Academy, Niels Bohr Institute, Blegdamsvej 17, 2100 Copenhagen, Denmark}

\date{\today}

\begin{abstract}
We study extreme-mass-ratio systems in theories admitting the Schwarzschild solution and propagating a massive graviton. We show that, in addition to small corrections to the quadrupolar and higher-order modes, a dipolar mode is excited in these theories and we quantify its excitation. While LIGO-Virgo-KAGRA observations are not expected to impose meaningful constraints in the dipolar sector, future observations by the Einstein Telescope or by LISA, together with bounds from dispersion relations, can {\it rule out} theories of massive gravity admitting vacuum General Relativistic backgrounds. For the bound to be circumvented, one needs to move away from Ricci-flat solutions, and enter a territory where constraints based on wave propagation and dispersion relations are not reliable.
\end{abstract}

\preprint{ET-0104A-23}

\maketitle

\section {Introduction.} 
There are compelling reasons to include massive degrees of freedom in the description of fundamental interactions.
%
To begin with, massive fields provide a framework to test a massless theory, for example by using observations to place upper bounds on the mass of the interaction carrier. In addition, the introduction of another scale in the theory can potentially be used to solve some of the outstanding problems, namely the dark matter and dark energy puzzles~\cite{Weinberg:1988cp,Barrow:2011zp,Hui:2016ltb}. Accordingly, a consistent modification of
Maxwell’s equations preserving the invariance of electro-dynamics under transformations of special relativity, yet endowing the photon with a mass, was considered by Proca in 1936~\cite{Proca:1,Proca:2,Proca:3,Goldhaber:2008xy}. 
Massive spin-2 fields were studied shortly afterwards by Fierz and Pauli~\cite{Fierz:1939ix}, and a nonlinear massive completion of General Relativity (GR) has been pursued ever since~\cite{deRham:2014zqa,deRham:2016nuf}. 

Bounds on the mass of the graviton can be obtained in a variety of ways. A graviton mass adds a Yukawa-like term to the strength of the gravitational interaction~\cite{Will:2018gku}, and either table-top experiments or the motion of planets in the solar system can be used to search for such deviations, within a mass range scaling inversely to the size of the laboratory~\cite{Bernus:2019rgl,Mariani:2023eka}. So far, bounds based on gravitational-wave (GW) emission belong to two categories. Superradiant-based bounds use the fact that Kerr black holes (BHs) are unstable and shed spin away in a two-step process (first condensing a cloud of gravitons in their exterior, and then eventually releasing all rotational energy as GWs)~\cite{Brito:2014wla,Brito:2015oca,Brito:2020lup,Dias:2023ynv}. Thus, observations of highly spinning massive BHs yield a constraint on the graviton mass $\mu \lesssim 5\times 10^{-23}\,{\rm eV}$~\cite{Brito:2013wya}. Perhaps the best-studied constraints are derived from dispersion relations as a GW propagates~\cite{Will:1997bb}. For massive gravitons, their propagation speed $v_g$ depends on their frequency $\omega$, as $v_g^2/c^2=
1-c^2/(\omega \bar{\lambda}_g)^2$, with $\bar{\lambda}_g=G/(\mu c^2)$ being the reduced graviton Compton wavelength. 
Dispersion changes the phase morphology of the GW as it propagates, producing changes with respect to predictions from GR. Current LIGO-Virgo-KAGRA (LVK) results yield the bound $\mu \lesssim 1.27\times 10^{-23}\, {\rm eV}$~\cite{LIGOScientific:2021sio}. Dispersion-relation-based bounds always assume that the waveform obtained in the local wave zone of the system is the same as that in (massless) GR, produced by the same BHs.

The full dynamical content of the field equations is not 
explored via dispersion relations. Here, we point out that 
the underlying {\it assumption} that sources are the same 
as in vacuum GR allows us to calculate rigorously GW 
generation effects and to uncover a dipolar mode 
which dominates emission at small graviton mass $\mu$. 
We now show that sources that resemble those
of GR (their structure and motion) lead to a massive
GW spectrum that can be ruled out by observations of future Earth- and space-based detectors.
Hereafter we use units such that the speed of light and 
Newton's constant $c=G=1$.

\section{Setup.} 
Propagation of massive spin-2 fields is strongly constrained by requiring the absence of ghosts~\cite{Fierz:1939ix}. The equations of motion governing massive spin-2 fluctuations $h_{\mu\nu}$ on a Ricci-flat background metric $g_{\mu\nu}$ are unique and read~\cite{Deser:1983mm,Higuchi:1986py,Bengtsson:1994vn,Porrati:2000cp,Bernard:2017tcg} 
\begin{equation}\label{e0}
G^{(1)}_{\mu\nu}[h]+\frac{\mu^2}{2}\left(h_{\mu\nu}-h g_{\mu\nu}\right)=8\pi T_{\mu\nu} \, ,
\end{equation}
where $G^{(1)}_{\mu\nu}[h]$ is the linearised Einstein tensor and $T_{\mu\nu}$ the energy-momentum tensor of matter. This means that any theory that admits Ricci-flat solutions and contains a massive graviton is governed by Eq.~\eqref{e0}. To circumvent Eq.~\eqref{e0}, one should allow for radical departures from standard vacuum GR, such as breaking Lorentz symmetry~\cite{DeFelice:2015hla,DeFelice:2022mcd}, including additional (dynamical) metrics \cite{Babichev:2014oua}, etc. We focus on the minimal extension of GR, and consider exclusively Eq.~\eqref{e0}. We also assume that the energy and momentum of matter sources is conserved, that is,
\begin{equation}\label{e1}
\nabla^{\mu} T_{\mu\nu}=0 \, .
\end{equation}
Then, from \eqref{e0} and \eqref{e1} it follows that $h_{\mu\nu}$ satisfies the constraints  
\begin{align}\label{e2}
\nabla^{\mu}h_{\mu\nu}-\nabla_{\nu}h=0,\ \ \  h=-\frac{16\pi}{3\mu^{2}}T \, ,
\end{align}
and the dynamical equation
\begin{align}\label{e3}
\left[\Delta_{L}+\mu^{2}\right]h_{\mu\nu}=S_{\mu\nu} \, ,
\end{align}
where $\Delta_{L}$ is the Lichnerowicz operator and $S_{\mu\nu}$ is determined by the source energy-momentum tensor
\begin{equation}\label{e4}
    S_{\mu\nu}\equiv 16\pi T_{\mu\nu}+\frac{16\pi}{3\mu^{2}}\nabla_{\mu}\nabla_{\nu}T-\frac{16\pi}{3}T g_{\mu\nu} \, .
\end{equation}
Some terms in \eqref{e2} and \eqref{e4} are singular as $\mu\to0$. For gauge theories of lower spin, this class of small-mass divergences are not present as long as the theory is coupled to conserved sources (e.g.~one recovers Maxwell's theory in taking the massless limit of a massive spin-1 vector or Proca field). Quite remarkably, it was established decades ago that the same is not true for spin-2 fields such as the graviton, where the coupling of conserved sources to some of the massive degrees of freedom (the helicity-0 mode in particular) persists in the massless limit. This is known as the van Dam, Veltman and Zakharov (vDVZ) discontinuity \cite{vanDam:1970vg,Zakharov:1970cc}, which in essence is the statement that GR is not recovered from the $\mu\to0$ limit of massive gravity. Arguments based on the so-called Vainshtein mechanism \cite{Vainshtein:1972sx} suggest that non-linearities of the massive theory could suppress the new degrees of freedom within certain scales \cite{deRham:2014zqa}. Therefore we expect that a viable theory of massive gravity displays a Vainshtein mechanism for stars (thereby circumventing solar system constraints). Nevertheless, such a theory can still contain BH solutions identical to those of GR, and known examples abound~\cite{Babichev:2015xha}. Such solutions are the subject of this work.

\section{Dipolar modes.} 
Linear massive gravity \eqref{e0}, unlike its massless 
counterpart, does not enjoy the gauge symmetry 
$h_{\mu\nu}\to h_{\mu\nu}+2\nabla_{(\mu}X_{\nu)}$. Thus, 
some degrees of freedom that are pure gauge in GR become 
physical and dynamical in the massive theory, and are excited in astrophysical scenarios. 
In particular, we now show that dipolar gravitational radiation can be
dominant for binaries, including Extreme Mass Ratio Inspirals 
(EMRI), which evolve in the milliHz LISA band 
\cite{LISA:2022yao}.

The equations governing the dynamics of massive fluctuations 
are most conveniently derived using a fully covariant approach, 
inspired by Refs.~\cite{Gerlach:1979rw,Gerlach:1980tx,Martel:2005ir,
Kodama:2003kk,Ishibashi:2011ws,Kodama:2003jz}. We shall write the Schwarzschild's metric in the general form
\begin{align}\notag
ds^{2}&=-f(r)dt^{2}+\frac{dr^{2}}{f(r)}+r^{2}\left(d\theta^{2}+\sin^{2}\theta d\phi^{2}\right)\\ \label{e5}
&=g_{ab}dy^{a}dy^{b}+r^{2}(y)\Omega_{AB}d\theta^{A}d\theta^{B} \, ,
\end{align}
where $f(r)=1-2M/r$, $y^{a}$ are any coordinates parametrising the ``$t-r$'' plane, $\theta^{A}$ parametrise the round 2-sphere with metric $\Omega_{AB}$, and it will be useful to introduce the radial vector $r_{a}\equiv (dr)_{a}$. Consider a metric fluctuation $h_{\mu\nu}$ of dipolar structure, that is, of the form
\begin{align}\notag
h&=p_{ab}(y)Y(\theta)dy^{a}dy^{b}+2q_{a}(y)Z_{A}(\theta)dy^{a}d\theta^{A}\\ \label{e6}
&+r^{2}(y) K(y) U_{AB}(\theta) d\theta^{A}d\theta^{B} \, ,
\end{align}
where $Y(\theta),Z_{A}(\theta),U_{AB}(\theta)$ are the dipolar ($l=1$) even harmonic tensors on the sphere \cite{Martel:2005ir} and $p_{ab}(y),q_{a}(y)$ and $K(y)$ are tensors in the space spanned by $y^{a}$ (the ``$t-r$'' plane). In GR such a mode is pure gauge, and consequently there is no dipolar radiation. In massive gravity, however, the lack of gauge symmetry makes \eqref{e6} physical, and its dynamical evolution, governed by \eqref{e2}-\eqref{e3}, leads to dipole emission. In terms of the variables
\begin{equation}\label{e7}
X\equiv r (r^{a}r^{b}p_{ab})\ ,\ Y\equiv r^{a}q_{a}\ , \ Z\equiv r K \ ,
\end{equation}
the equations of motion \eqref{e2}-\eqref{e3} for the mode \eqref{e6} can be reduced (as explained in Appendix \ref{ApDipole}) to the system of equations
\begin{equation}\label{sys1}
\left(\square-\bold{V}\right)\left(\begin{array}{@{}c@{}} X \\ Y \\ Z \end{array}\right)=\left(\begin{array}{@{}c@{}} \Sigma_{X} \\ \Sigma_{Y} \\ \Sigma_{Z} \end{array}\right) \, ,
\end{equation}
where $\square$ denotes the d'Alembertian of $g_{ab}$, the matrix potential reads 
\begin{align}
\bold{V}=\left(\begin{array}{@{}ccc@{}}\mu ^2-\frac{10 M}{r^3}+\frac{6}{r^2} &\frac{24 M-8 r}{r^3}& -\frac{4 \left(15 M^2-9 M r+r^2\right)}{r^4} \\ -\frac{2}{r^2} &\mu ^2-\frac{16 M}{r^3}+\frac{6}{r^2} &\frac{2 (r-3 M)}{r^3}  \\ -\frac{2}{r^2}&\frac{4}{r^2} & \mu ^2-\frac{10 M}{r^3}+\frac{4}{r^2}  \end{array}\right)
\end{align}
and $\Sigma_{X},\Sigma_{Y},\Sigma_{Z}$ are terms associated to the source, whose general expression for arbitrary $T_{\mu\nu}$ is reported in Appendix \ref{ApDipole}. In the case that the source is a point particle, these scale as $\Sigma_{X,Y,Z}=(1/\mu^{2})\tilde{\Sigma}_{X,Y,Z}$, with $\tilde{\Sigma}_{X,Y,Z}$ being regular as $\mu\to0$ (see \eqref{scaling}). Thus, the variables $(\tilde{X},\tilde{Y},\tilde{Z})\equiv\mu^{2}(X,Y,Z)$ satisfy a system of partial differential equations (PDEs) that can be solved numerically and are unproblematic as $\mu\to0$.

The new massive degrees of freedom contribute to the 
power radiated during astrophysical processes, like 
EMRI coalescences. 
In flat space, far from the sources, there is a well-defined 
notion of energy-momentum tensor of a massive spin-2 field, 
which reads \cite{PhysRevD.65.044022,Cardoso:2018zhm}
\begin{equation}\label{emTens}
t_{\mu\nu}=\frac{1}{32\pi}\langle \nabla_{\mu}h_{\alpha\beta}  \nabla_{\nu}h^{\alpha\beta} - \nabla_{\mu}h \nabla_{\nu}h\rangle \, , 
\end{equation}
and in terms of this, the power radiated at infinity is
\begin{equation}\label{flux}
\dot{E}=\frac{dE}{dt}=-\int_{S} t^{\mu}r^{\nu}t_{\mu\nu}d\Omega_S \, ,
\end{equation}
where $t^{\mu}=\left(\partial_{t}\right)^{\mu}$, and $S$ denotes 
a very distant spherical shell that encloses all sources. A dipolar wave with frequency $\Omega$ is described by $(X,Y,Z)=e^{-i\Omega t}(X(r),Y(r),Z(r))$, and evaluating  
\eqref{flux} on-shell gives
%
%
\begin{equation}\label{fluxmono}
\dot{E}=\frac{\sqrt{\Omega^{2}-\mu^{2}}}{16\pi \lvert \Omega\rvert}\left[2\mu^{2}\lvert Y_{\infty} \rvert^{2}+3\Omega^2\lvert Z_{\infty} \rvert^{2}\right] \, ,
\end{equation}
where $Y_{\infty},Z_{\infty}$ are the asymptotic values 
of $Y(r)$ and $Z(r)$ (more details can be found in Appendix \ref{ApDipole}). Equation \eqref{fluxmono} holds 
so long $\mu<\lvert\Omega\rvert$. For graviton masses 
$\mu>\lvert\Omega\rvert$ the energy emission of 
the massive modes is exponentially suppressed.
The excitation of $Y_{\infty}$ and $Z_{\infty}$ depends on the details of the source.

Finally, it is illustrative to explore the effect of 
dipole waves on nearby geodesics at infinity. Consider 
an inertial observer at distance $d$ from a source, and 
associate to its worldline a parallely-propagated frame 
\cite{Misner:1973prb}. In such a frame, and to leading order in $1/d$, the 
components of the geodesic deviation vector $(S_{1},S_{2})$ transverse to the direction of wave propagation of a dipole mode read
\begin{equation}\label{gedev}
    \left(\begin{array}{@{}c@{}} S_{1} \\ S_{2} \end{array}\right)= \left(\begin{array}{@{}c@{}} S^{(0)}_{1}-\frac{\vert Y_{\infty}\vert}{2d}\left(\frac{\mu}{\Omega}\right)^{2}\sqrt{\frac{3}{2 \pi }} \left[S_{3}^{(0)}   \cos \left( \Omega \tau -\varphi_{0} \right)\right]  \\ S^{(0)}_{2}-\frac{\vert Y_{\infty}\vert}{2d}\left(\frac{\mu}{\Omega}\right)^{2}\sqrt{\frac{3}{2 \pi }} \left[S_{3}^{(0)} \sin \left( \Omega \tau -\varphi_{0} \right)\right] \end{array}\right)
\end{equation}
where $S^{(0)}_{i},\varphi_{0}$ are constants defined in Appendix \ref{ApDipole}, and $\tau$ is the proper time of the inertial observer. The relative motion of free-falling test particles immersed on a GW is, therefore, different from the usual one GR. In particular, a free-falling observer would see that free-falling test particles move in circles warping the direction of wave propagation. This motion, in addition, exhibits some longitudinal oscillation, even though there is no relative time dilation. We notice that a massive spin-2 wave would, in general, exhibit more polarisations than those described in \eqref{gedev}, but here we focus on the waves that are excited by an EMRI. See Appendix \ref{ApDipole} for more details.

\section{Dipole radiation from EMRIs.} 
We now focus on astrophysical scenarios 
provided by EMRIs, in which the source terms in 
Eq.~\eqref{sys1} describe a
point particle of mass $m_{p}$ in circular motion around a Schwarzschild BH, with orbital radius $r_{p}$. 
For the dipole $m=1$ mode the GW frequency is 
then fixed by the geodesic equation to be 
$\Omega_{p}=\left(M r_{p}^{-3}\right)^{1/2}$, and the source 
terms in \eqref{sys1} read
\begin{equation}\label{scaling}
   \left(\begin{array}{@{}c@{}} \Sigma_{X} \\ \Sigma_{Y} \\ \Sigma_{Z} \end{array}\right)=e^{-i\Omega_{p} t}\frac{m_{p}/M^{2}}{r_{p}^{2}\mu^{2}}\left(\begin{array}{@{}c@{}} \tilde{\Sigma}_{X}(r) \\ \tilde{\Sigma}_{Y}(r) \\ \tilde{\Sigma}_{Z}(r) \end{array}\right) \, ,
\end{equation}
where $\tilde{\Sigma}_{X,Y,Z}(r)$ are dimensionless distributions that depend smoothly on $\mu$, and whose explicit form can be found in the Appendix \ref{app:source_circular}.
We solved this problem numerically with two independent 
codes, one in the frequency and the other in the time 
domain, based on methods employed in Refs.~\cite{Sundararajan:2007jg, Macedo:2013jja, Cardoso:2022whc}. In both codes, the 
point particle is approximated by a smoothed distribution 
$\delta \left(r-r_p\right) = \exp[-(r-r_p)^2/(2\sigma^2)]/(\sqrt{2\pi}\sigma)$, where the value of $\sigma$ is chosen to guarantee 
numerical convergence of the solution. 
The two codes agree within the numerical error when varying 
$\sigma$, the extraction radius of the fields, and the location 
of the BH horizon. Our results confirm and extend those of Ref.~\cite{Cardoso:2018zhm}, by pushing them to the limits with two independent codes.

\begin{figure}[t]
    \centering
    \includegraphics[width=\columnwidth]{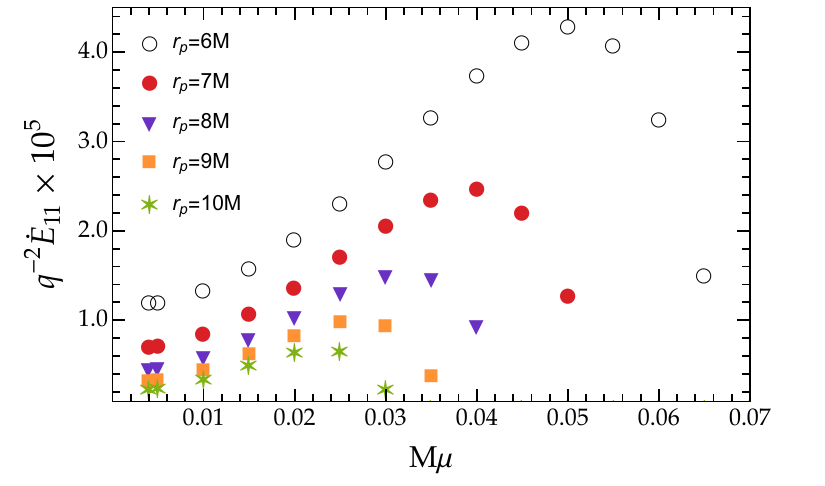}
    \caption{Energy flux emitted in the dominant dipolar mode $l=m=1$ as a function of the graviton mass $M\mu$, for different radius $r_p$ of the particle in circular orbit. The mass ratio $q\equiv m_p/M$. For small $M\mu$, the flux follows a dipolar behavior given by Eq.~\eqref{flux_dipolar_small_mu}.
    }%
    \label{fig:Flux}
\end{figure}
Our findings are summarized in Fig.~\ref{fig:Flux}, showing the 
energy flux carried by the dipolar mode as a function of the 
graviton mass $M\mu$ for different orbital radii $r_p$. For fixed $r_p$
the energy flux peaks around $M \mu \sim (M/r_p)^{3/2}$ and vanishes for $\mu>\lvert\Omega\rvert$. Formally, the wavefunctions $X,Y,Z \sim e^{i \sqrt{\Omega^2-\mu^2} r}$ at large distances, so large $\mu$ fluctuations are not propagating waves but are exponentially suppressed instead. For $\mu<\lvert\Omega\rvert$ the flux is nonzero, and the dipolar sector is excited. 

The numerical solution is challenging to obtain at very small $\mu$ (see Eq.~\eqref{scaling}): the source term diverges as $\mu \rightarrow 0$, so one needs to work with high arithmetic precision. In addition to that, the convergence of the asymptotic values of the field $Y_\infty \, , Z_\infty$ with the extraction radius becomes slower. 
Nonetheless, our results are consistent with a finite flux in the $\mu \rightarrow 0$ limit. Numerically, we find that $Y_{\infty}\mu$ goes to zero while $Z_\infty$ is the quantity that contributes to the non-zero flux at graviton masses. We obtain a behavior consistent with a dipolar scaling given by
\begin{equation}
\dot{E}^{l=1} \underset{\mu \rightarrow 0}{\approx} 10^{-2} \frac{q^2 M^4}{r_p^4}\,,\label{flux_dipolar_small_mu}
\end{equation}
where the mass ratio $q\equiv  m_p/M$. This is only an approximation to our results, valid for $\mu\ll \Omega$ and $r_p\gg M$. In other words, the $\mu\to 0$ limit gives rise to important dipolar radiation, but to negligible dispersion as the graviton propagates. This is a crucial point of our results.

\section{Bounds on the graviton mass.} 
For small $M\mu$, and including the dominant quadrupolar term $\dot{E}_{\rm N}=32/5 \,q^2 (M/r_p)^5$, we can express the total luminosity as ($\Theta$ is the Heaviside function)
\beq
\dot{E}&=&\dot{E}_{\rm N}\left(1+B\frac{r_p}{M}\right)\,,\\
%
B&=&2\times 10^{-3}\Theta(\Omega-\mu)\,.\label{flux_correction}
\eeq
There are a few interesting aspects of this result. The first is that the dipolar contribution competes with the next-to-leading order correction to the Newtonian result, which takes the form $\dot{E}=\dot{E}_{\rm N}(1-1247\Omega^2 r_p^2/336)$~\cite{Fujita:2012cm}. Already for $r_p\approx 40 M$ the dipolar term in massive gravity is of the same order and {\it opposite} sign.

Expression \eqref{flux_correction} is in the form used by Ref.~\cite{Barausse:2016eii} to bound dipolar emission. LISA can bound $B\lesssim 10^{-5}$ or better for EMRIs, thus being able to exclude practically all the interesting region of parameter space $\mu <\vert \Omega\rvert$. From Fig.~\ref{fig:Flux} a rough estimate is then that LISA-type instruments would {\it exclude} $M\mu \lesssim 0.03,0.01$ for systems which enter the LISA band at $r_p=10M,20M$, respectively. In other words, the constraint for EMRIs is of order $\mu\gtrsim 10^{-15}\,\frac{1\,5\, M_\odot}{M}\, {\rm eV}$. Current LVK constraints, however, imply that $\mu \lesssim 1.27\times 10^{-23}\, {\rm eV}$~\cite{LIGOScientific:2021sio}, and are based on the same assumption (that BHs belong to the vacuum GR family), together with dispersion-relation bounds. In other words, LISA has the ability to exclude massive gravity altogether, or then force one to go beyond the Ricci-flat paradigm when studying massive gravity from a GW perspective. Note that the central BH for an EMRI is expected to be spinning, while our results only describe non-spinning geometries, but we don't expect any qualitative difference with respect to the above.

\begin{figure}[t]
    \centering
    \includegraphics[width=\columnwidth]{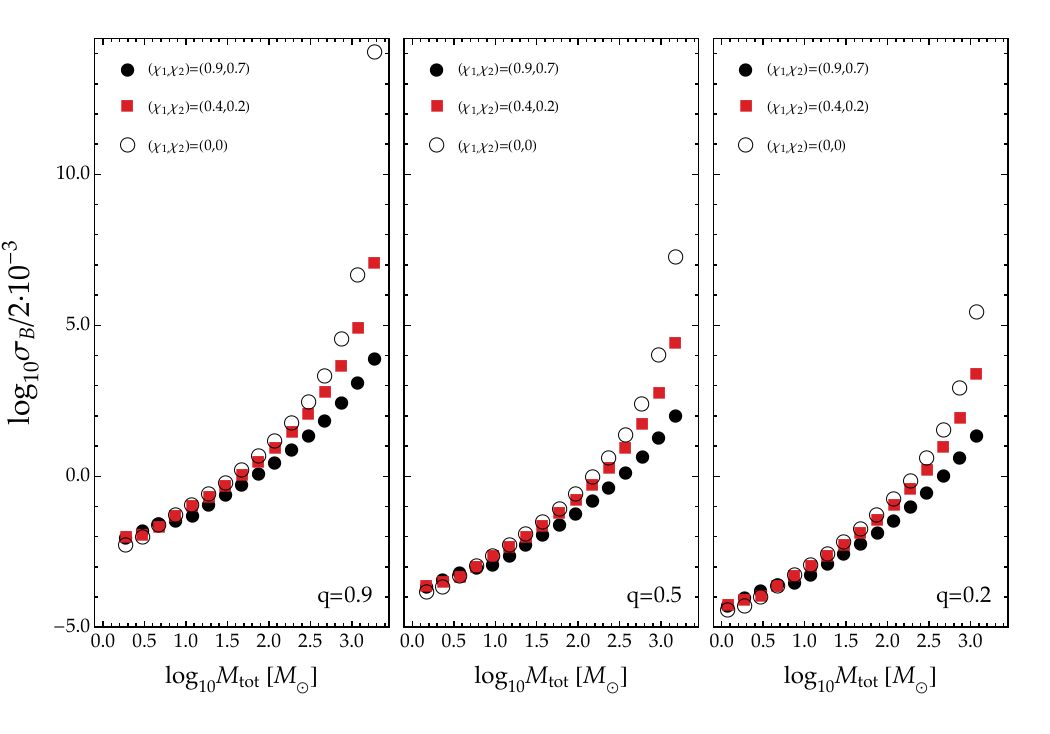}
    \caption{1-$\sigma$ uncertainty on the dipole 
    parameter $B$ inferred by ET observations of BH 
    binaries with total detector-frame mass 
    $M_\textnormal{tot}$, at $d=500$Mpc from the 
    detector, assuming average orientation. 
    Left, centre and right panels show 
    results for mass ratio $q=0.9,0.5$ and 0.2, 
    while colored dots represent binaries 
    with different spin $\chi_{1,2}$ configurations. 
    For all calculations we assumed two 
    aligned L-shaped Einstein Telescope detectors in their ET-D configuration 
    \cite{Hild:2010id}.}%
    \label{fig:sigma}
\end{figure}
The extrapolation of our results to nearly equal-mass systems requires moving away from the regime of validity of result~\eqref{flux_correction}. In GR the extrapolation yields sensible results when the mass ratio $q$ is promoted to a symmetric mass ratio, $q\to q/(1+q)^2$~\cite{Berti:2007fi,LeTiec:2014oez}. For dipolar contributions, a geometric correction term $1-q$ appears when computing the source, see Appendix \ref{app:source_circular} for details. Thus, we expect 
$\dot{E}=\dot{E}_{\rm N}\left(1+B(1-q)^2\frac{r_p}{M}\right)$.
%
The extrapolation is not unique, and full numerical relativity simulations are required in this regime, but we can take it as a rough guide. LVK can constraint $B (1-q)^2\lesssim 10^{-3}$ for stellar-like BHs~\cite{Barausse:2016eii} and is thus below the threshold of placing constraints for GW150914-like systems ($q$ is too close to unity). Note also that LVK constrains $B\lesssim 10^{-5}$ from the GW170817 event~\cite{LIGOScientific:2018dkp}, but it involves neutron stars and therefore outside of our (vacuum) framework. 
However, the Einstein Telescope (ET) promises 
far better forecasts. Figure~\ref{fig:sigma} shows 
the constraints on $B$ that can be inferred by 
ET through GW observations of comparable mass 
sources on circular orbits with different  
configurations. We compute errors using a 
Fisher matrix approach \cite{Vallisneri:2007ev,Favata:2010ic}, 
adopting for the GW signals a TaylorF2 
waveform model describing the inspiral phase of 
the binary in the frequency domain 
\cite{Damour:2000gg, Arun:2004hn, Buonanno:2009zt} (technical details on the error calculations are discussed in the Appendix \ref{app:parestimation}).
The uncertainties strongly 
depend on the mass components, which determine 
the timescale of the binary evolution in the 
detector band. While errors on $B$ 
deteriorate as $q\rightarrow 1$, 
we expect for $q\lesssim 0.5$ and 
$M_\textnormal{tot}\lesssim 100M_\odot$ that ET 
can determine $B\lesssim 10^{-4}$, placing 
constraints on massive gravity competitive 
with those inferred by LISA.

\section { Discussion.} 
Dipolar emission of GWs in theories of massive gravity is a compelling mechanism to bound the graviton mass or to exclude the theory altogether. We find, with a fully relativistic analysis, that dipolar emission is so strong that it can possibly rule out theories of massive gravity. To realize completely the potential of this analysis, a proper handling of extreme mass ratio systems needs to be obtained. Our conclusions are based on the analysis of Ref.~\cite{Barausse:2016eii}, but EMRIs are complex systems and their understanding is far from being under control. Likewise, our results suggest that Earth-based experiments may impose equally impressive constraints, but a proper data analysis with the full inspiral-merger-ringdown should be studied.

We {\it assume} that $T_{\mu\nu}$ is conserved, but it suffices that $\nabla^{\mu}T_{\mu\nu}\to0$ as $\mu\to0$ \cite{deRham:2014zqa}. However, steering away from this condition would also change the motion of point particles, invalidating dispersion-relation-based bounds. Similarly, it can be argued that the perturbative approach in powers of mass ratio $q$ may not be well defined at very small $\mu$ (a strong coupling problem). The structure of higher order terms depends on the particular nonlinear completion of the theory. Nevertheless, i) our bounds refer to masses $M\mu$ which can be much larger than $q$, leading one to suspect that for some theories there is a well-defined perturbative hierarchy. ii) failure to converge at small $M\mu$ would again imply that wave generation cannot be assumed to occur as in GR, invalidating also previous LVK bounds. 
A main goal in our work is to show that sources that resemble those of GR (their structure and motion) lead to a massive GW spectrum that is incompatible with observations. 
It should be noted also that in some massive gravity theories, Schwarzschild BHs are afflicted by a linear instability mechanism~\cite{Brito:2013wya,Babichev:2013una,Gregory:1993vy}. Nevertheless, the instability acts on a spherically symmetric mode and is expected to play no important role in the dynamics of the dipolar mode. In fact, its nonlinear development -- in a specific theory -- leads to a hairy BH, which can be made arbitrarily close to a Schwarzschild BH, where backreaction is never important~\cite{Brito:2013wya,Brito:2013xaa,Gervalle:2020mfr}. We don't expect this mechanism to change in any relevant way the bounds we derived for the dipolar mode. In addition, the instability is long-wavelength in nature and its time scale is pushed to large values for small enough graviton mass. Dipolar quasinormal modes and bound states of BHs in such theories were computed in Ref~\cite{Brito:2013wya}, but their excitation amplitude is yet to be calculated.

Our results are very general, and constrain {\it any} theory of massive gravity admitting a Schwarzschild (and possibly Kerr) background. Arbitrarily small graviton masses are not allowed as they would lead to GW emission that can be ruled out by observations. There are two important lessons which should be learned: a. the dynamical content of the theory -- beyond simple dispersion relations -- is important. We find that most of the emission takes place in a dipolar mode which is absent in vacuum GR, and which requires the relativistic calculation of fluxes and metric perturbations. b) constraints based on dispersion relations are oblivious to wave generation: they assume that GWs are generated ``as in vacuum General Relativity'' and then change the propagation properties of the waves. However, we show that the assumption that the background is the same as General Relativity leads to a non-perturbative behavior at small graviton masses and consequent strong lower bounds on its mass. To evade the bounds we establish here, one needs to change completely the background and/or the inertial motion of fields. Thus, wave generation will radically depart from vacuum General Relativity, rendering an analysis on dispersion relation invalid. In summary, GW-based constraints on the mass of the graviton should be based on the full dynamical equations rather than just on dispersion relation arguments. Our results highlight the need to perform full numerical relativity simulations of theories beyond GR.

\noindent {\bf \em Acknowledgments.} 
We are indebted to Evgeny Babichev, Richard Brito, Gregorio Carullo, Cristiano Germani, Shinji Mukohyama and Paolo Pani for critical comments and discussions.
V.C.\ is a Villum Investigator and a DNRF Chair, supported by VILLUM Foundation (grant no. VIL37766) and the DNRF Chair program (grant no. DNRF162) by the Danish National Research Foundation. V.C. acknowledges financial support provided under the European Union’s H2020 ERC Advanced Grant “Black holes: gravitational engines of discovery” grant agreement
no. Gravitas–101052587. Views and opinions expressed are however those of the author only and do not necessarily reflect those of the European Union or the European Research Council. Neither the European Union nor the granting authority can be held responsible for them.
This work was supported by FCT under project No. 2022.01324.PTDC.
F. D. acknowledges financial support provided by FCT/Portugal through grant No. SFRH/BD/143657/2019.
This project has received funding from the European Union's Horizon 2020 research and innovation programme under the Marie Sklodowska-Curie grant agreement No 101007855 and No 101007855.

\bibliography{ref}

\clearpage

\appendix

\onecolumngrid
\section{Dipolar modes}\label{ApDipole}
This section contains the additional details about dipolar modes that are referred to in the main text. First, we consider aspects related to the master wave equations, and then discuss in more detail the asymptotic energy flux and geodesic deviation by dipolar waves. 

\subsection*{Equations of Motion}
A dipolar (even) gravitational mode is described by a metric fluctuation of the form\footnote{Odd dipolar modes are not relevant for the class of sources we are concerned with in this work.}
\begin{equation}\label{dip}
h=p_{ab}Y(\theta)dy^{a}dy^{b}+2q_{a}Z_{A}(\theta)dy^{a}d\theta^{A}+r^{2} K U_{AB}(\theta)d\theta^{A}d\theta^{B}
\end{equation}
where $p_{ab},q_{a},K$ are tensors in the ``$t-r$'' plane (in arbitrary coordinates) and $Y(\theta),Z_{A}(\theta),U_{AB}(\theta)$ are the usual (even) tensor harmonics with harmonic number $l=1$ \cite{Martel:2005ir}. Plugging \eqref{dip} into the equations of motion Eqs.~\eqref{e2}-\eqref{e3}, and using the orthogonality properties of tensor harmonics \cite{Martel:2005ir}, the angluar dependence is factored out and the equations become a system of coupled linear PDEs on the ``$t-r$'' plane. More precisely, Eq.~\eqref{e2} gives the following first-order PDEs on the ``$t-r$'' plane,
\begin{align}\label{l1c1}
 \Sigma &=p^{a}_{a}+2K\,,\\\label{l1c2}
 0&=r^{-2}D_{a}\left(r^{2}p^{a}_{b}\right)-\frac{2}{r}D_{b}\left(rK\right)-D_{b}p^{c}_{c}-\frac{2}{r^{2}}q_{b}\,,\\\label{l1c3}
0&=r^{-2}D_{a}\left(r^{2}q^{a}\right)-K-p^{c}_{c}\, ,
\end{align}
where $D_{a}$ is the covariant derivative of $g_{ab}$, while the dynamical equation Eq.~\eqref{e3} gives (we recall that $f=1-2M/r$)
\begin{align}\label{ab2}
\Sigma_{ab}=&-\square p_{ab}-\frac{2}{r}D^{d}r D_{d}p_{ab} +\frac{4}{r^{2}}p_{c(b}D_{a)}rD^{c}r+\left(\frac{2}{r^{2}}-f''+\mu^{2}\right)p_{ab}+f'' p^{c}_{c} g_{ab}\\\notag
&-\frac{8}{r^{3}}q_{(a}D_{b)}r+\left(2\frac{f'}{r}g_{ab}-\frac{4}{r^{2}} D_{a}rD_{b}r\right)K\, , \\ \label{a2}
\Sigma_{a}=&-\square q_{a}+\left(\frac{f+1}{r^2}+\mu^{2}\right)q_{a}+\frac{4}{r^2}q_{b}D^{b}rD_{a}r-\frac{2}{r}p_{ab}D^{b}r+2\frac{D_{a}r}{r}K\, , \\\label{L2}
\Sigma_{L}=&-D^{a}\left(r^{2}D_{a}K\right)+\left(4f+\mu^{2}r^{2}\right)K+\frac{4}{r}q_{a}D^{a}r-2p_{ab} D^{a}rD^{b}r +rf' p^{a}_{a}\, ,
\end{align}
where a prime denotes a radial derivative, and the source terms $\Sigma,\Sigma_{ab},\Sigma_{a},\Sigma_{L}$ are given by
\begin{align}
\Sigma\equiv&-\frac{16\pi}{3\mu^{2}}\int d\Omega \, \bar{Y}T\,,\ \ \Sigma_{ab}\equiv\int d\Omega\, \bar{Y} S_{ab}\,, \ \ \Sigma_{a}\equiv\frac{1}{2}\int d\Omega\,\bar{Z}_{A}\Omega^{AB}S_{aB}\,,\ \ \Sigma_{L}\equiv\frac{1}{2}\int d\Omega \, \bar{U}_{AB}S_{CD}\Omega^{AC}\Omega^{BD}\,,
\end{align}
where $T$ is the trace of the energy-momentum tensor $T_{\mu\nu}$, and $S_{\mu\nu}$ is given by Eq.~\eqref{e4}.
%
%
To obtain the wave equations Eq.~\eqref{sys1} for the metric variables $X,Y,Z$ introduced in Eq.~\eqref{e7}, consider the following combinations of source terms,
\begin{align}
\Sigma_{X}&\equiv-r\left(r^{a}r^{b}\Sigma_{ab}\right)+2rf'r^{a}D_{a}\Sigma+r\left[(f f')'-\frac{1}{2}(f')^{2}\right]\Sigma\, ,\\
\Sigma_{Y}&\equiv-r^{a}\Sigma_{a}+f'\Sigma \, ,\\
\Sigma_{Z}&\equiv-r^{-1}\Sigma_{L}+f'\Sigma \, .
\end{align}
Then, using the equations of motion to write $\Sigma_{ab},\Sigma_{a},\Sigma$ and $\Sigma_{L}$ in terms of the metric fluctuation and its derivatives (that is, the right hand sides of \eqref{l1c1}-\eqref{L2}) one obtains precisely the system of equations Eq.~\eqref{sys1}, relating the source terms $\Sigma_{X},\Sigma_{Y}$ and $\Sigma_{Z}$ to the metric fluctuations $X,Y,Z$.

The above derivation is valid for any conserved energy-momentum tensor $T_{\mu\nu}$. Here we are interested in a point particle source of mass $m_{p}$ which follows a circular geodesic of radius $r_{p}$ at the equator. The specific energy $E$, angular momentum $L$ per unit rest mass and orbital frequency $\Omega_{p}$ read
\begin{equation}
    E=\left(1-\frac{3M}{r_{p}}\right)^{-1/2}\left(1-\frac{2M}{r_{p}}\right)\, , \ \ L=\left(1-\frac{3M}{r_{p}}\right)^{-1/2}\left(M r_{p}\right)^{1/2}\, , \ \ \Omega_{p}= (M r_{p}^{-3})^{1/2} \, ,
\end{equation}
while the particle's four-velocity is $u_{\mu}=(-E,0,0,L)$ and its energy-momentum tensor
reads
\begin{align}\notag
T^{\mu\nu}=\frac{m_p f(r_{p})}{r_{p}^{2}E}u^{\mu}u^{\nu}\delta(r-r_{p})\delta(\theta-\pi/2)\delta(\phi-\Omega_{p} t)\, .
\end{align}
Using this energy-momentum tensor the source terms $(\Sigma_{X},\Sigma_{Y},\Sigma_{Z})$ can be evaluated straightforwardly and reduce to
\begin{align}
\Sigma_{X}&=e^{-i\Omega_{p} t}\frac{m_p}{r_{p}^{2}\mu^{2}}\sqrt{\frac{2\pi (r_{p}-3M)}{3r_{p}}}\left[\mathcal{X}_{0} \delta(r-r_{p})+\mathcal{X}_{1}(r) \delta'(r-r_{p})+\mathcal{X}_{2}(r) \delta''(r-r_{p})\right]\, ,\\
\Sigma_{Y}&=e^{-i\Omega_{p} t}\frac{m_p}{r_{p}^{2}\mu^{2}}\sqrt{\frac{2\pi (r_{p}-3M)}{3r_{p}}}\left[\mathcal{Y}_{0} \delta(r-r_{p})+\mathcal{Y}_{1}(r) \delta'(r-r_{p})\right]\, ,\\
\Sigma_{Z}&=e^{-i\Omega_{p} t}\frac{m_p}{r_{p}^{2}\mu^{2}}\sqrt{\frac{2\pi (r_{p}-3M)}{3r_{p}}}\left[\mathcal{Z}_{0} \delta(r-r_{p})+\mathcal{Z}_{1}(r) \delta'(r-r_{p})\right]\, ,
\end{align}
where
\begin{align}
\mathcal{X}_{0}&=-\frac{40 M^2}{r_{p}^3}-8 \mu ^2 M+\frac{16 M}{r_{p}^2}+4 \mu ^2 r_{p}\, , \ \ \mathcal{X}_{1}(r)=\frac{20 M (2 M-r)}{r^2} \, , \ \ \mathcal{X}_{2}(r)=-\frac{4 (r-2 M)^2}{r}\, ,\\
\mathcal{Y}_{0}&=\frac{4 (r_{p}-4 M)}{r_{p}^2}\, , \ \ \mathcal{Y}_{1}(r)=\frac{8 M}{r}-4\, ,\\
\mathcal{Z}_{0}&=-\frac{8 M}{r_{p}^2}-\frac{6 \mu ^2 M r_{p}}{3 M-r_{p}}+4 \mu ^2 r_{p}+\frac{4}{r_{p}}\, , \ \ \mathcal{Z}_{1}(r)=\frac{8 M}{r}-4\, .
\end{align}
\subsection*{Energy flux and geodesic deviation at infinity}
Far from the sources, the solution for $(X,Y,Z)$ of Eqs.~\eqref{sys1} describing outgoing waves reads
\begin{align}
X&=e^{-i\Omega_{p} t+i r \sqrt{\Omega_{p}^{2}-\mu^{2}}-i\frac{ M \left(\mu ^2-2 \Omega_{p} ^2\right)}{\sqrt{\Omega_{p} ^2-\mu ^2}}\log{(r/2M)}}\left(X_{\infty}+\sum_{i=1}^{\infty}X_{i}r^{-i}\right)\, ,\\
Y&=e^{-i\Omega_{p} t+i r \sqrt{\Omega_{p}^{2}-\mu^{2}}-i\frac{M \left(\mu ^2-2 \Omega_{p} ^2\right)}{\sqrt{\Omega_{p} ^2-\mu ^2}}\log{(r/2M)}}\left(Y_{\infty}+\sum_{i=1}^{\infty}Y_{i}r^{-i}\right)\, ,\\
Z&=e^{-i\Omega_{p} t+i r \sqrt{\Omega_{p}^{2}-\mu^{2}}-i\frac{ M \left(\mu ^2-2 \Omega_{p} ^2\right)}{\sqrt{\Omega_{p} ^2-\mu ^2}}\log{(r/2M)}}\left(Z_{\infty}+\sum_{i=1}^{\infty}Z_{i}r^{-i}\right)\, ,
\end{align}
where the only free parameters are the amplitudes $(X_{\infty},Y_{\infty},Z_{\infty})$, while the coefficients $(X_{i},Y_{i},Z_{i})$ are fixed in terms of the latter by requiring that Eqs.~\eqref{sys1} hold order by order. Imposing the remaining equations of motion Eqs.~\eqref{e2}-\eqref{e3} fixes $X_{\infty}=-\frac{2\Omega_{p}^{2}}{\mu^{2}}Z_{\infty}$, while $Y_{\infty}$ and $Z_{\infty}$ remain independent. Finally, plugging this solution into the energy-flux formula Eq.~\eqref{flux} gives the expression in Eq.~\eqref{fluxmono}. 

It is also possible to compute the geodesic deviation induced by such dipolar waves. Let $(t,x,y,z)$ be almost inertial coordinates at some large distance $d$ from the source and for simplicity fix their origin at the axis defined by $\partial_{\phi}$ (the fixed points of $\partial_{\phi}$), so the orbital plane of the particle is perpendicular to the observer's line of sight that points towards the source. We can always align the $z$-axis with the (positive sense of the) axis of $\partial_{\phi}$, so that the $(x,y)$-plane is ``parallel'' to the orbital plane. In these coordinates, and to leading order in $1/d$ the wave is described by
\begin{equation}\label{assymMetric}
h_{\mu\nu}=\frac{Y_{\infty}}{2 d}\sqrt{\frac{3}{2 \pi }}\left(
\begin{array}{cccc}
 0 & \frac{\sqrt{\Omega_{p} ^2-\mu ^2}}{\Omega_{p} } & \frac{i \sqrt{\Omega_{p} ^2-\mu ^2}}{\Omega_{p} } & 0 \\
 \frac{\sqrt{\Omega_{p} ^2-\mu ^2}}{\Omega_{p} } & 0 & 0 & -1 \\
 \frac{i \sqrt{\Omega_{p} ^2-\mu ^2}}{\Omega_{p} } & 0 & 0 & -i \\
 0 & -1 & -i & 0 \\
\end{array}
\right)e^{i\left(-\Omega_{p} t+z\sqrt{\Omega_{p}^{2}-\mu^{2}}+d\sqrt{\Omega_{p}^{2}-\mu^{2}}\right)}\, .
\end{equation}
As one would expect, we obtain a wave that propagates outwards along the $z$-axis. It induces some rotation in the transverse $(x,y)$-plane, while it is polarised in the $(x,z)$ and $(y,z)$ planes. The amplitude that governs the leading order in $1/d$ (i.e. the plane wave behaviour) is $Y_{\infty}$, while $Z_{\infty}$ is relevant only for subleading terms $O(d^{-2})$. However, both $Y_{\infty}$ and $Z_{\infty}$ contribute to the energy flux per unit of solid angle (see Eq.~\eqref{fluxmono} of the main text). Now we can use (the real part of) \eqref{assymMetric} to obtain the effect of the wave on nearby geodesics. Consider a free-falling observer with four-velocity $u^{\mu}$, and let $e^{\mu}_{A}=(e^{\mu}_{0},e^{\mu}_{1},e^{\mu}_{2},e^{\mu}_{3})=(u^{\mu},e^{\mu}_{i})$ be a parallely-propagated orthonormal frame along the observer's worldline, that is, it satisfies
\begin{equation}
 u^{\mu}\nabla_{\mu}e_{A}^{\nu}=0\, ,\ \ u_{\mu}e^{\mu}_{i}=0\, ,\ \ e_{i\mu}e^{\mu}_{j}=\delta_{ij}\,.
\end{equation}
Then, the geodesic deviation equation reads \cite{Misner:1973prb}
\begin{equation}\label{devgeo}
\frac{d^{2} S_{A}}{d\tau^{2}}=R_{A00B}S^{B}\ ,
\end{equation}
where $\tau$ is the observer's proper time and $S^{A}$ and $R_{ABCD}$ are the components of the (infinitesimal) deviation vector and Riemann tensor relative to the frame $e_{A}$. Assume now that $u^{\mu}$ is initially at the origin of our almost inertial coordinates $(t,x,y,z)$ at infinity introduced above. Then, $e_{A}\approx(\partial_{t},\partial_{x},\partial_{y},\partial_{z})$ and $\tau\approx t$, so one can solve \eqref{devgeo} by working perturbatively on the amplitude $Y_{\infty}$. Choosing the integration constants such that the nearby geodesics would be at rest (relative to each other) when there is no wave ($Y_{\infty}=0$), we obtain the solution
\begin{align}
S_{0}&=0\, ,\\
S_{1}&=S^{(0)}_{1}-\frac{\vert Y_{\infty}\vert}{2d}\left(\frac{\mu}{\Omega_{p}}\right)^{2}\sqrt{\frac{3}{2 \pi }} \left[S_{3}^{(0)}   \cos \left( \Omega_{p} \tau -\varphi_{0} \right)\right] \, ,\\
S_{2}&=S^{(0)}_{2}-\frac{\vert Y_{\infty}\vert}{2d}\left(\frac{\mu}{\Omega_{p}}\right)^{2}\sqrt{\frac{3}{2 \pi }} \left[S_{3}^{(0)} \sin \left( \Omega_{p} \tau -\varphi_{0} \right)\right]\, ,\\
S_{3}&=S_{3}^{(0)}-\frac{\vert Y_{\infty}\vert}{2d}\left(\frac{\mu}{\Omega_{p}}\right)^{2}\sqrt{\frac{3}{2 \pi }}\left[S_{1}^{(0)}\cos \left( \Omega_{p} \tau -\varphi_{0} \right)+S^{(0)}_{2}\sin \left( \Omega_{p} \tau -\varphi_{0} \right)\right]\, ,
\end{align}
where the phase $\varphi_{0}$ is given by
\begin{equation}
\varphi_{0}=\text{Arg}\left(Y_{\infty}\right)+d\sqrt{\Omega_{p}^{2}-\mu^{2}}\, .
\end{equation}
That is, the free-falling observer would see that close by free-falling test particles move in circles warping the direction of wave propagation. This motion, in addition, exhibits some longitudinal oscillation, too, even though there is no relative time dilation, in the sense that $S_{0}=0$ along the observer's worldline. 

\section{Source terms for circular Newtonian orbits\label{app:source_circular}}
Here we work out the source terms not for a black hole background, but for two pointlike particles in a Minkowski background. We wish to show that the dipolar source vanishes for equal-mass systems and that it scales as $1-q$ for $q\sim 1$ (hence fluxes scale as $(1-q)^2$ in this regime). The energy-momentum tensor in the non-relativistic limit reads
\begin{equation}
T_{\mu\nu}=\left[\frac{m_{1}}{r_{1}^{2}}\delta(r-r_{1})\delta(\theta-\pi/2)\delta(\phi-\Omega_{0}t)+\frac{m_{2}}{r_{2}^{2}}\delta(r-r_{2})\delta(\theta-\pi/2)\delta(\phi+\pi-\Omega_{0}t)\right]\delta^{t}_{\mu}\delta^{t}_{\nu}
\end{equation}
where
\begin{equation}
r_{1}=\frac{m_{2}}{m_{1}+m_{2}}r_{0},\ \ \ \ r_{2}=\frac{m_{1}}{m_{1}+m_{2}}r_{0}, \ \ \ \ \Omega_{0}=\sqrt{m_{1}+m_{2}}r_{0}^{-3/2}.
\end{equation}
Going through the definitions, we find the source terms
\begin{align}
\Sigma_{X}=&e^{-i t \Omega }4 \sqrt{\frac{2 \pi }{3}}(m_{1}+m_{2})\left\{X_{0,1}\delta\left(r-r_{1}\right)+X_{0,2}\delta\left(r-r_{2}\right)+X_{2,1}(r)\delta''\left(r-r_{1}\right)+X_{2,2}(r)\delta''\left(r-r_{2}\right)\right\}\\
\Sigma_{Y}=&e^{-i t \Omega }4 \sqrt{\frac{2 \pi }{3}}(m_{1}+m_{2})\left\{Y_{0,1}\delta\left(r-r_{1}\right)+Y_{0,2}\delta\left(r-r_{2}\right)+Y_{1,1}\delta'\left(r-r_{1}\right)+Y_{1,2}\delta'\left(r-r_{2}\right)\right\}\\
\Sigma_{Z}=&e^{-i t \Omega }4 \sqrt{\frac{2 \pi }{3}}(m_{1}+m_{2})\left\{Z_{0,1}\delta\left(r-r_{1}\right)+Z_{0,2}\delta\left(r-r_{2}\right)+Z_{1,1}\delta'\left(r-r_{1}\right)+Z_{1,2}\delta'\left(r-r_{2}\right)\right\}
\end{align}
where
\begin{align}
X_{0,1}&=-\frac{ m_{2}  }{m_{1} r_{0}}\, ,\ \ \ X_{2,1}(r)=\frac{ m_{2} r (m_{1}+m_{2}) }{\mu ^2 m_{1}^2 r_{0}^2}\, ,\\ \notag \\
Y_{0,1}&=-\frac{ m_{2} (m_{1}+m_{2})^2 }{\mu ^2 m_{1}^3 r_{0}^3}\,  , \ \ \ Y_{1,1}=\frac{ m_{2} (m_{1}+m_{2}) }{\mu ^2 m_{1}^2 r_{0}^2}\,,\\ \notag \\
Z_{0,1}&=-\frac{ m_{2}   \left(m_{1}^2 \mu ^2 r_{0}^2+(m_{1}+m_{2})^{2}\right) }{\mu ^2 m_{1}^3 r_{0}^3}\, , \ \ \ Z_{1,1}=\frac{ m_{2} (m_{1}+m_{2}) }{\mu ^2 m_{1}^2 r_{0}^2}\, ,
\end{align}
while $(X_{i,2},Y_{i,2},Z_{i,2})=(-X_{i,1}\left(m_{1}\leftrightarrow m_{2}\right),-Y_{i,1}\left(m_{1}\leftrightarrow m_{2}\right),-Z_{i,1}\left(m_{1}\leftrightarrow m_{2}\right))$. It is now easy to see that indeed the source terms can be combined when $m_1\sim m_2$ and scale like $\propto (1-q)$ as advertised.

\section{Parameter estimation with the Einstein Telescope\label{app:parestimation}}

We provide here technical details on the parameter 
estimation performed to compute bounds on the dipolar 
amplitude $B$, shown in Fig.~2 of the main text. 
We consider binary BH events observed by two L-shaped 
Einstein Telescope detectors, aligned with respect to 
each-other. We adopt the design ET-D sensitivity 
curve~\cite{Hild:2010id} for the interferometer noise 
spectral density.

We model the GW signal emitted by the binary using a 
TaylorF2 waveform model which describes the inspiral 
evolution of the coalescence. In the frequency 
domain, the GW signal is given by:
\be
\tilde h (f) = C_\Omega{\cal A}_\textnormal{PN}  e^{i \psi_\text{\tiny PP} (f) + i \psi_\text{\tiny ppE} (f)} \ . 
\ee
The waveform phase is described by the post-Newtonian (pN) 
expansion. In particular $\psi_\text{\tiny PP}$ contains 
terms up to the 3.5PN order~\cite{Damour:2000gg, Arun:2004hn, Buonanno:2009zt}, and depends on: (i) the binary chirp mass 
$\mathcal{M} = (m_1 m_2)^{3/5}/(m_1+m_2)^{1/5}$, (ii) 
the symmetric mass ratio $\eta = m_1 m_2/(m_1+m_2)^2$, 
with $m_{1,2}$ being the binary component masses, 
(iii) linear spin terms up to 3PN order through the 
(anti)symmetric combinations of the individual spin 
components $\chi_{s}=(\chi_{1}+\chi_{2})/2$ and 
$\chi_{a}=(\chi_{1}-\chi_{2})/2$, and quadratic spin 
corrections entering at 2PN order. Modification of the 
GR baseline due to massive gravity are included 
through the parametrized post-Einsteinian (ppE) phase 
\begin{equation}\label{math:pper}
\psi_\textnormal{ppE}=-\frac{3}{224} \eta^{2/5} (1-q)^2
B  ({\cal M} \pi f)^{-7/3}
\end{equation}
where $q=m_1/m_2$ is the binary mass ratio \cite{Barausse:2016eii}. 
We assume the leading Newtonian term for the amplitude 
\be
\mathcal{A}_\textnormal{PN} = \sqrt{\frac{5}{24}} 
\frac{\mathcal{M}^{5/6}f^{-7/6}}{\pi^{2/3}d_L}\ ,
\ee
where $d_L$ is the luminosity distance. The geometric 
factor $C_{\Omega}$ depends on the source position in the 
sky, and on its orientation with respect to the detector. 
We assume here average orientation, such that $C_\Omega=2/5$.
The overall waveform model depends on 7 parameters 
$\vec{\theta}=\{{\cal M},\eta,\chi_s,\chi_a,t_c,\phi_c,B\}$, 
where $(t_c,\phi_c)$ are the time and phase at the coalescence, 
which we both fix to zero.

We study the detectability of the parameter $B$ using a 
Fisher-matrix approach~\cite{Vallisneri:2007ev}, in which the posterior distribution of 
$\vec{\theta}$ can be described by a multivariate Gaussian 
distribution centered around the {\it true} values 
$\vec{\hat{\theta}}$, with covariance ${\bf \Sigma} = {\bf \Gamma}^{-1}$, where
\be
\Gamma_{ij}= \left\langle \frac{\partial h}{\partial \theta_i}\bigg\vert\frac{\partial h}{\partial \theta_j}\right\rangle_{\vec{\theta}=\vec{\hat{\theta}}}
\ee
is the Fisher information matrix, 
and we have introduced the scalar product over the 
detector noise spectral density $S_n(f)$ between two waveforms 
$h_{1,2}$ as:
\be
\langle h_1\vert 
h_2\rangle=4\Re\int_{f_\text{\tiny min}}^{f_\text{\tiny max}} \frac{\tilde{h}_1(f)\tilde{h}^\star_2(f)}{S_n(f)}df \ ,\label{scalprod}
\ee
where $\star$ denotes complex conjugation. 
The integral \eqref{scalprod} is performed assuming 
$f_\textnormal{min}=3$Hz, while $f_\textnormal{max}$ is 
given by the ISCO frequency for the Kerr metric 
including self-force corrections \cite{Favata:2010ic}. 
We have varied the maximum frequency to assess the 
stability of our calculations, and computed errors 
on the parameter $B$ scaling $f_\textnormal{max}\rightarrow f_\textnormal{max}/2$. Overall we 
find very small changes with respect to the results 
discussed in the main text, but for the largest 
masses we analyse, namely for $M_\textnormal{tot}\gtrsim 300M_\odot$.

The fisher approach provides a reliable approximation of 
the real posterior distribution  for signals with large 
signal-to-noise ratio, as those expected for the Einstein 
Telescope. With the Fisher Matrix in hand, the statistical 
error on the $i$-th parameter is given by the diagonal 
component $\sigma_i=\Sigma^{1/2}_{ii}$. 

As a final remark we note that, while the TaylorF2 approximant is able 
to capture the (early) inspiral evolution of a binary coalescence, it is not suited to describe the late stages before the 
merger. To bridge this gap more sophisticated waveform models need to get informed from numerical relativity simulations, and perturbation theory, to provide a full description of the merger and ringdown 
phases \cite{Khan:2015jqa}. Different choices for 
$\tilde{h}(f)$ would affect the forecasts on the parameter's errors. However, we expect such choice to not change dramatically bounds we infer on $B$. Indeed the pre-Newtonian dipole term introduced 
in Eq.~\eqref{math:pper} modifies the waveform in a 
low frequency range where the TaylorF2 approximant is 
indistinguishable from other models 
\cite{Yunes:2016jcc}. 

Finally, we note that the use of the Fischer information matrix calls for large signal to noise ratios, and there are subtleties that should be taken into account when establishing precise bounds~\cite{Vallisneri:2007ev}.

\end{document}